\documentclass[12pt]{article}

\usepackage[margin=2.5cm]{geometry}

\usepackage{amsmath,amsthm,amsfonts,amscd,amssymb,bbm,mathrsfs,enumerate,url}
\usepackage{csquotes}

\usepackage{graphicx,tikz}
\usepackage[pdfborder={0 0 0}]{hyperref}
\usetikzlibrary{matrix,arrows}
\usetikzlibrary{decorations.pathreplacing}
\usetikzlibrary{decorations.pathmorphing}
\usetikzlibrary{patterns,fadings}

\def\RR{{\mathbb R}}

\def\1{{\mathbbm 1}}

\def\diff{{\rm Diff}}

\def\diffs1{\diff(S^1)}

\def\psl2r{{\rm PSL}(2,\RR)}
\def\sl2r{{\rm SL}(2,\RR)}
\def\su11{{\rm SU}(1,1)}
\def\2dmob{{\overline{\psl2r}\times\overline{\psl2r}}}
\def\<{\langle}
\def\>{\rangle}

\newif\ifLetterStyle

\LetterStylefalse

\title{Further comments on
``Individual external dose monitoring of all citizens of Date {C}ity by
  passive dosimeter 5 to 51 months after the {F}ukushima {NPP} accident (series):
  1. {C}omparison of individual dose with ambient dose rate monitored by aircraft
  surveys'': Inconsistencies in Table 1 2014 Q3 and Figure 4f
}
\date{} 

\newcommand{\myauthors}{
{\bf Shin-ichi Kurokawa} \\
The High Energy Accelerator Research Organization (KEK), Tsukuba \\
email: {\tt shin-ichi.kurokawa@kek.jp}\\
{\bf Yutaka Hamaoka} \\
Faculty of Business and Commerce, Keio University\\
email: {\tt hamaoka@fbc.keio.ac.jp}\\
{\bf Kyo Kageura} \\
Graduate School of Education, University of Tokyo\\
email: {\tt kyo@p.u-tokyo.ac.jp}\\
{\bf Jun Makino} \\
Department of Planetology, Graduate School of
  Science, Kobe University\\
email: {\tt makino@mail.jmlab.jp}\\
{\bf Masaki Oshikawa} \\
Institute for Solid State Physics, University of Tokyo\\
email: {\tt oshikawa@issp.u-tokyo.ac.jp}\\
{\bf Yoh Tanimoto}
\\
   Dipartimento di Matematica, Universit\`a di Roma ``Tor Vergata''\\
   email: {\tt hoyt@mat.uniroma2.it}
}

\ifLetterStyle
\author{}
\else 
\author{\myauthors}
\fi 

\begin{document}
\maketitle
%

\ifLetterStyle
\vspace{0.3cm}
\noindent
Dear Sir,

\vspace{0.2cm}
\fi

In this letter, we point out serious inconsistencies in the paper~\cite{MH16}, in particular of Table 1 and Figure 4f.
Throughout this Letter, "Fig.\! XX" or "Table XX" refer to those in Ref.~\cite{MH16} unless specified otherwise.

\begin{enumerate}[{(}1{)}]

\item In Table 1, authors write in the box of Participants column that corresponds to the period 2014 Q2, Q3, Q4 and 2015 Q1 ``1. Ages $0\sim 15$; 2. Pregnant women; 3. Zone A \& B; 4. Zone C random sampling + applicant''.  This description is inconsistent with a magazine published by Date City: On \cite[p.2]{DateFSN26} it is written ``from July 2014 to June 2015, the number of participants who have continuously participated in the measurements is $12,912$; the participants of the period are ages $0\sim 15$, pregnant women, Zone A, the other Zones random sampling and applicant.'', hence not all residents in Zone B participated.  It should also be noted that the description, ``1. Ages $0 \sim 15$; 2. Pregnant women; 3. Zone A \& B; 4. Zone C random sampling + applicant'', is consistent for the period 2013 Q2, Q3, Q4 and 2014 Q1 with \cite[p.2]{DateFSN22}.

Moreover, in Section 2.4 of the paper, it is written "After that, all children up to the age of 18(sic), all pregnant women, and all persons living in zone A continued to participate in the measurement, while the number of participants living in zone B and C has been gradually reduced; in these zones, glass badges are distributed based on random sampling of the population, and also to residents who request one, instead of targeting the entire population."  \cite{MH16}.
In fact, the number of the participants changed not gradually but abruptly when random sampling started for zone C in July 2013 and again it started also for zone B in July 2014.     

\item\label{4fnumber} From July 2014, in addition to Zone C, the range of participants in Zone B had been changed from all citizens to randomly sampled citizens.  This change is supposed to have caused a substantial decrease of the number of participants of this period.
\begin{itemize}
\item In Table 1, six age distribution histograms of the glass-badge monitoring participants are shown.  Above the histogram that corresponds to the period 2014 Q3, it is written ``2014 3Q $\mathrm{N}=21080$'' \footnote{Incidentally, the authors of the paper use the two ways to show the period, such as Q3 and 3Q. We use the former style in this Letter.}. 
\item In Figure 4f that corresponds to the period October 2014 to December 2014 as well as the 9th airborne dose monitoring, on the one hand, it is written that $\mathrm{n}=21052$, where $\mathrm{n}$ means the number of the participants in this period.  This number is very close to that shown in Table 1 2014 Q3, namely $21,080$.  By adding the number of the participants of all of bins of the histogram shown below the box-and-whisker plot of this period, we obtain about $21,000$, which is also very close to $21,052$. 
\item In the document \cite{Date15RPMM}, on the other hand, it is reported that the number of glass-badges distributed to citizens of Date City for the period from October 2014 to December 2014 (2014 Q3) is $16,037$ and the percentage for these citizens to return back the glass-badges is about $90\%$. 
This suggests that the number of the participants is about $14,500$.
\end{itemize}
The inconsistency between $16,037$ (the number of glass-badges distributed
in the period 2014 Q3) and about $21,000$ (the number of participants in Table 1, 2014 Q3:\,$21,080$ and Figure 4f:\,$21,052$) is glaring. This entails that the histogram of
Table 1 2014 Q3 and Figure 4f are most likely not made by using the
correct data of the period 2014 Q3.

\begin{figure}[ht]
 \centering
 \includegraphics[width=12cm]{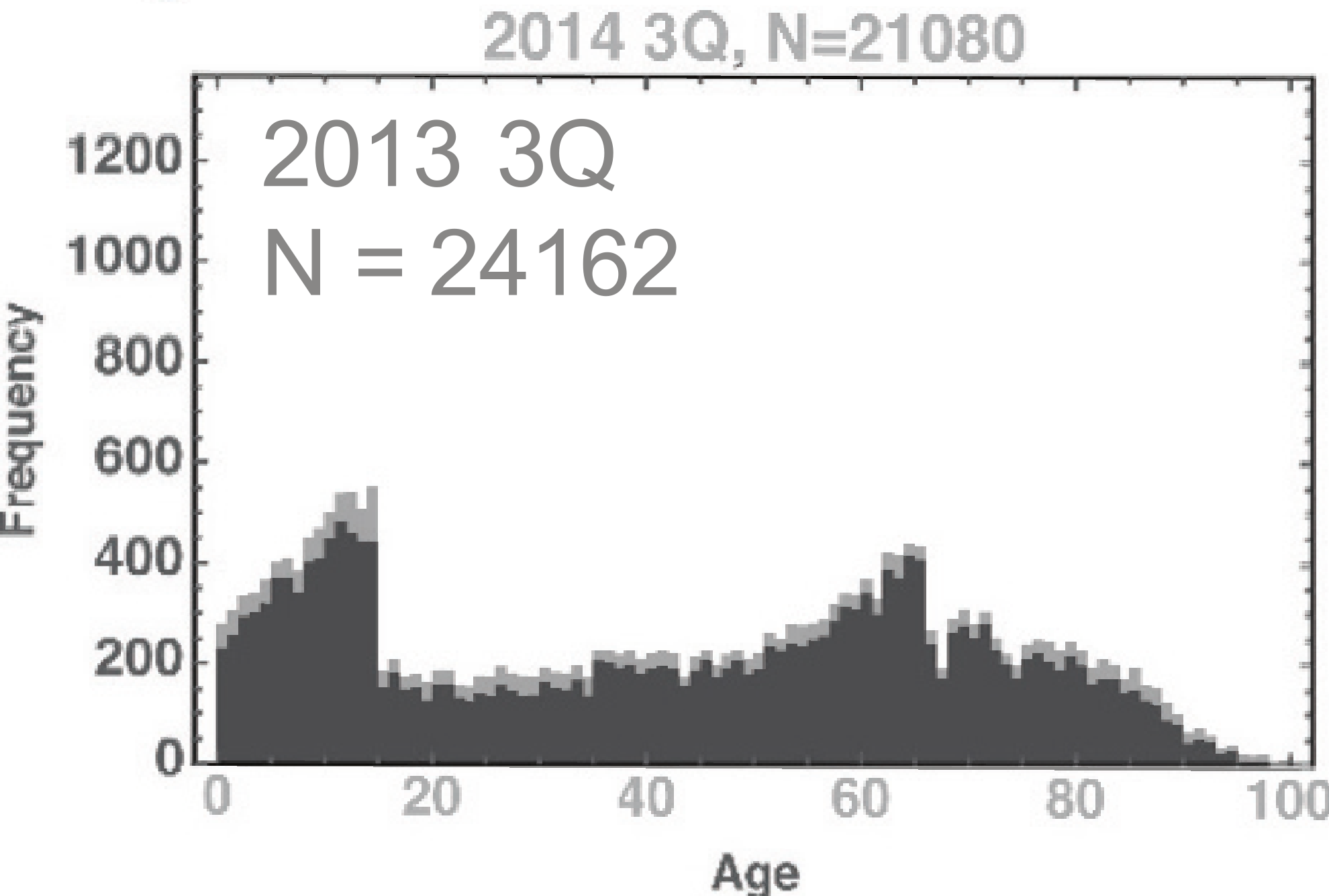}\\
 \includegraphics[width=12cm]{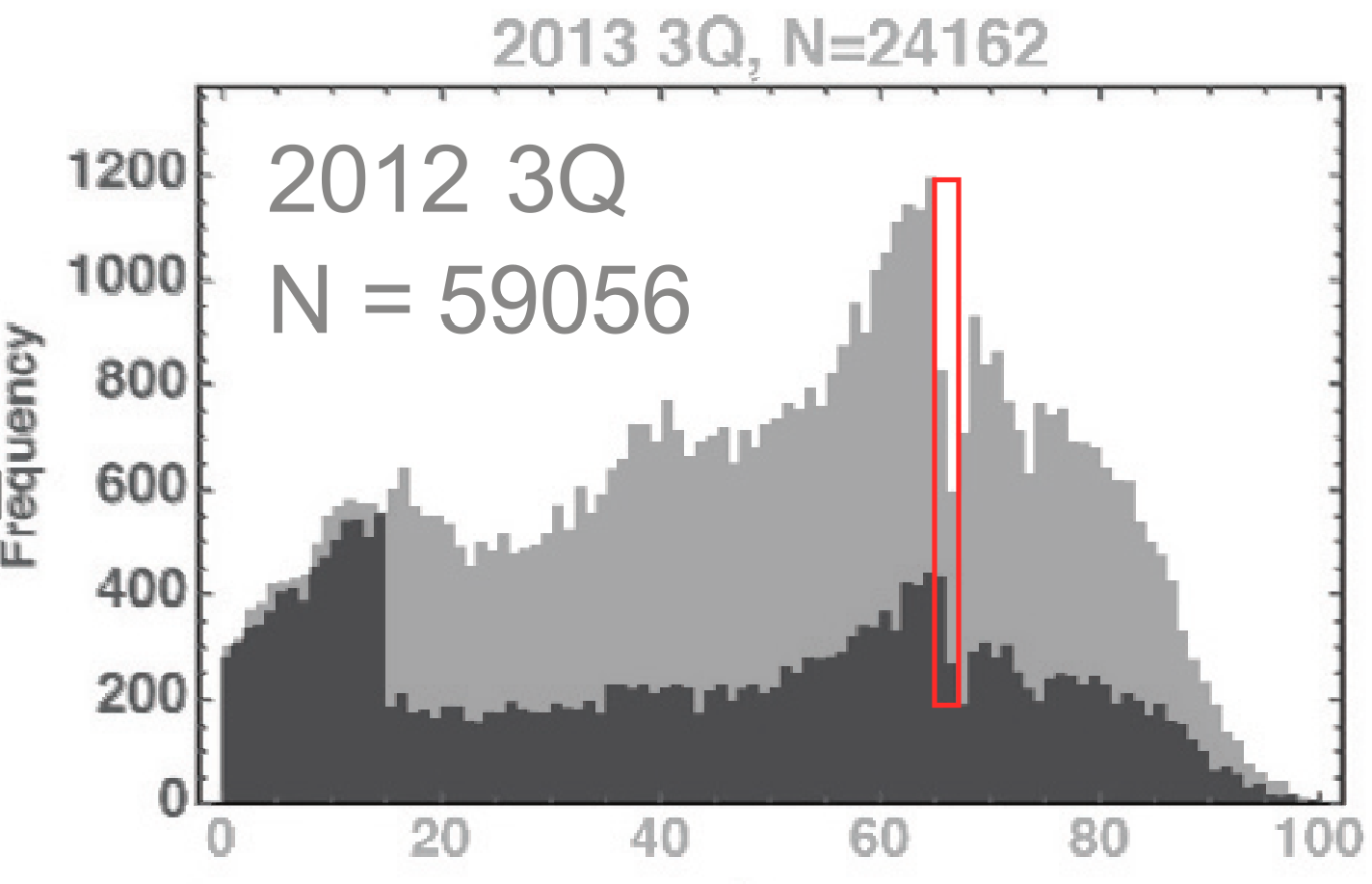}
    \caption{Comparison of the age distributions between 2013 Q3 and 2014 Q3 of Table 1 (top panel) and that between 2012 Q3 and 2013 Q3 (bottom panel).
    \textbf{ }
    }
  \label{fig:age-dist_2013Q3vs2014Q3}
\end{figure}

\item In top panel of Figure~\ref{fig:age-dist_2013Q3vs2014Q3} of this Letter, the age distribution histograms in Table 1 are compared by an overlay,
between 2013 Q3 and 2014 Q3.
We see that the age distribution is nearly identical, including peaks and dips.
However, this cannot be right on two respects:
\begin{itemize}
    \item First, as the correct number of the participants in 2014 Q3 is about $14,500$ as we discussed in (\ref{4fnumber}), the age distribution of the histogram should reflect this correct number of the participants.
    \item More importantly, because one year has passed between 2013 Q3 and 2014 Q3, the age of each individual must have increased by one.
However, the comparison shows that the significant peaks and dips exist at the same ages between 2013 Q3 and 2014 Q3, see top panel of Figure \ref{fig:age-dist_2013Q3vs2014Q3} of this Letter.
In particular, the significant dip at the age 67 in 2013 Q3 corresponds to the low number of birth around the end of the
Second World War (1945). This should appear at the age 68 in 2014 Q3 but remains at 67 in the histogram in Table 1. We also note that there is indeed a shift of the age distribution by one year between 2012 Q3 and 2013 Q3, as is shown in bottom panel of Fig. 1, hence the ages are defined at each measurement.
\end{itemize}
Both of these facts imply again that the data in the lowest histogram of Table 1 entitled ``2014 3Q'' do not correspond to the participants in the period 2014 Q3.

\begin{figure}
    \centering
      \includegraphics[width=12cm]{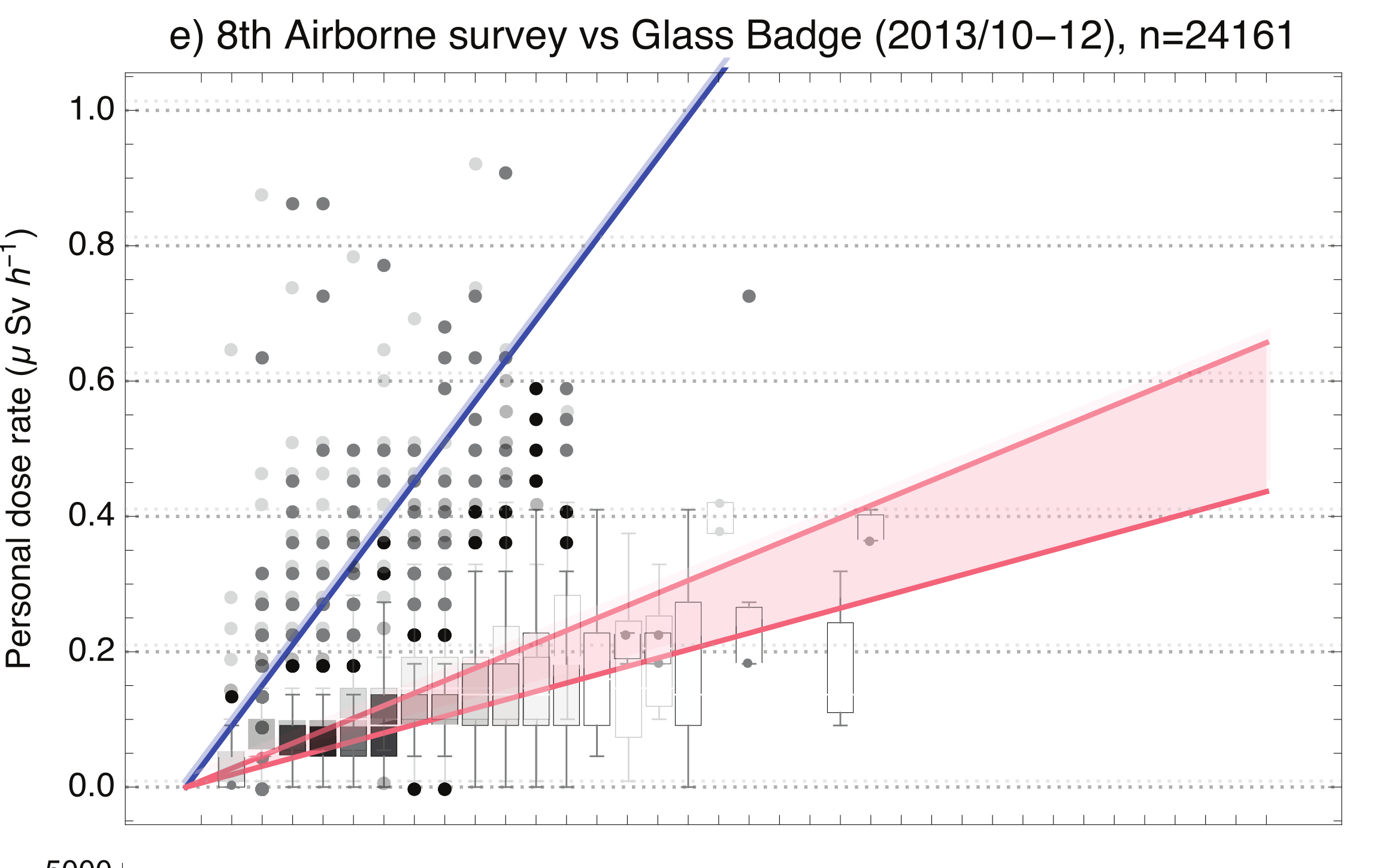}\\
\includegraphics[width=12cm]{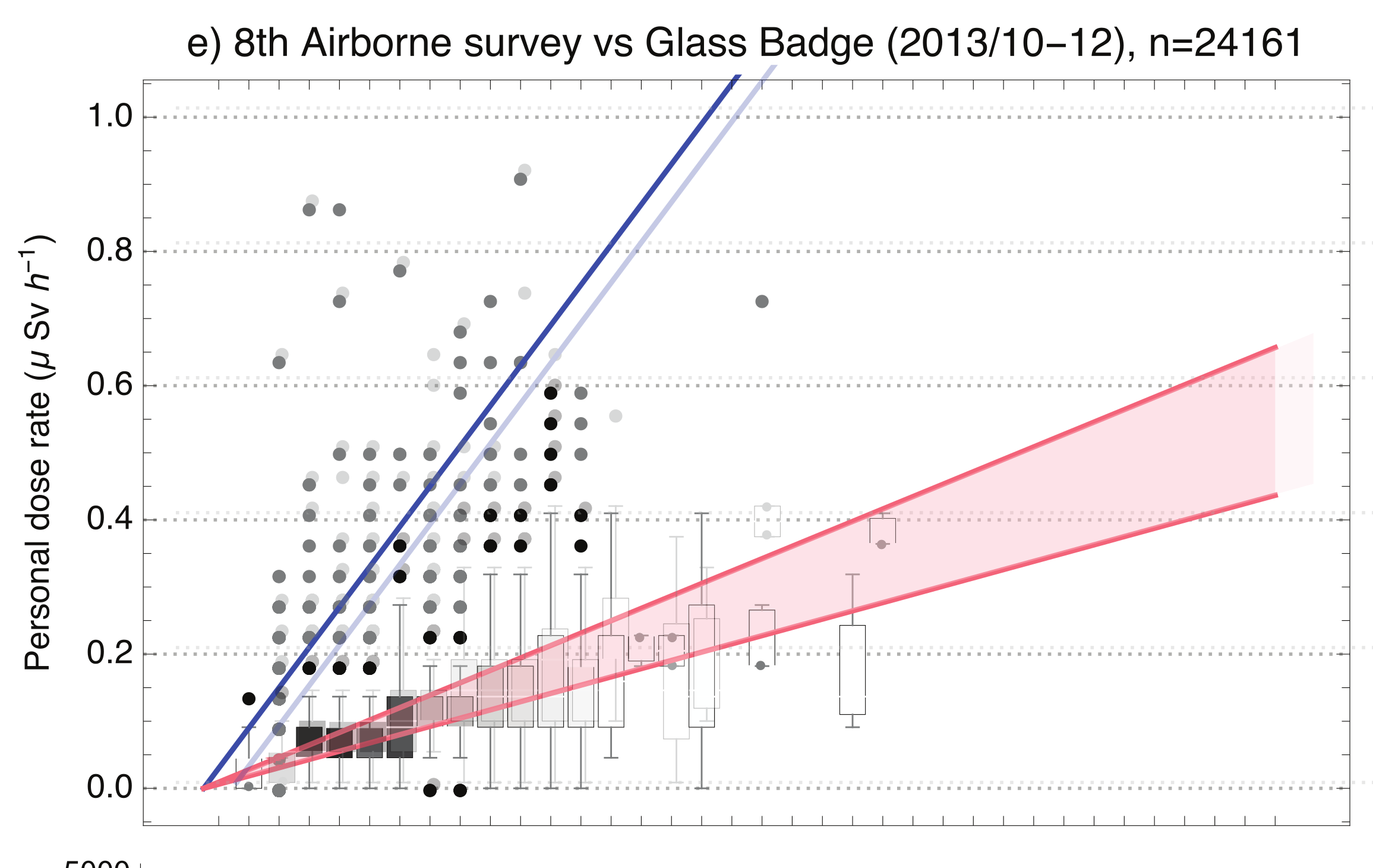}
    \caption{Comparison of dose distributions for 2013 Q3 and 2014 Q3.
    Top panel: two plots overlaid. Bottom panel: 2014 Q3 data shifted
    right by one bin. 
}
  \label{fig:fig4ef1overlay}
\end{figure}

\item\label{4e4f} In Figure~\ref{fig:fig4ef1overlay} of this Letter, we compare the dose distributions in Figs.\! 4e and 4f, which are for
2013 Q3 and 2014 Q3.
Even though they are supposed to be dose data obtained by glass-badge measurements in different periods, the two plots are strikingly similar.
In particular, after shifting the dose data by one bin (Fig.\! 4f to the right),
many points representing ``outliers'' overlap precisely.
It is highly unlikely that they are accidental agreements.

\item The observations (\ref{4fnumber})--(\ref{4e4f}) strongly suggest that 
the glass-badge data for 2014 Q3 were actually
a part of those for 2013 Q3, for some unknown reason, plotted against the ambient dose rate of 2014 Q3.
This can explain all the inconsistencies above: the wrong number of participants, the similarity between age distributions and the similarity between box-and-whisker plots up to the shift to the left due to the decrease of the ambient dose rate.

\item Each panel in Fig.\! 4 contains two red straight lines which seem to indicate the confidence interval of the coefficient
for the linear relation estimated from the data.
The authors write that the red lines show $c=0.12$ and $c=0.18$, respectively (although the ``confidence interval" is not clearly defined, see \cite{OHKKMT20}).
We found that, while the red lines meet at the origin in Figs.\! 4a--4e, they do not meet at the origin in Fig.\! 4f,
as it is seen clearly in Figure~\ref{fig:fig4ef-origin} of this Letter.
We also note that the font size of "0.0" is different between Figs.\! 4a--4e and Fig.\! 4f. If the authors had performed the same analysis and used the same software/script to produce Figs.\! 4a--4e (as they should),
there would be no reason to expect such a difference.
This suggest that not only the lines in Fig.\! 4f but also Fig.\! 4f itself are produced in an unusual and erroneous way. 

\begin{figure}
    \begin{center}
      \includegraphics[width=12cm]{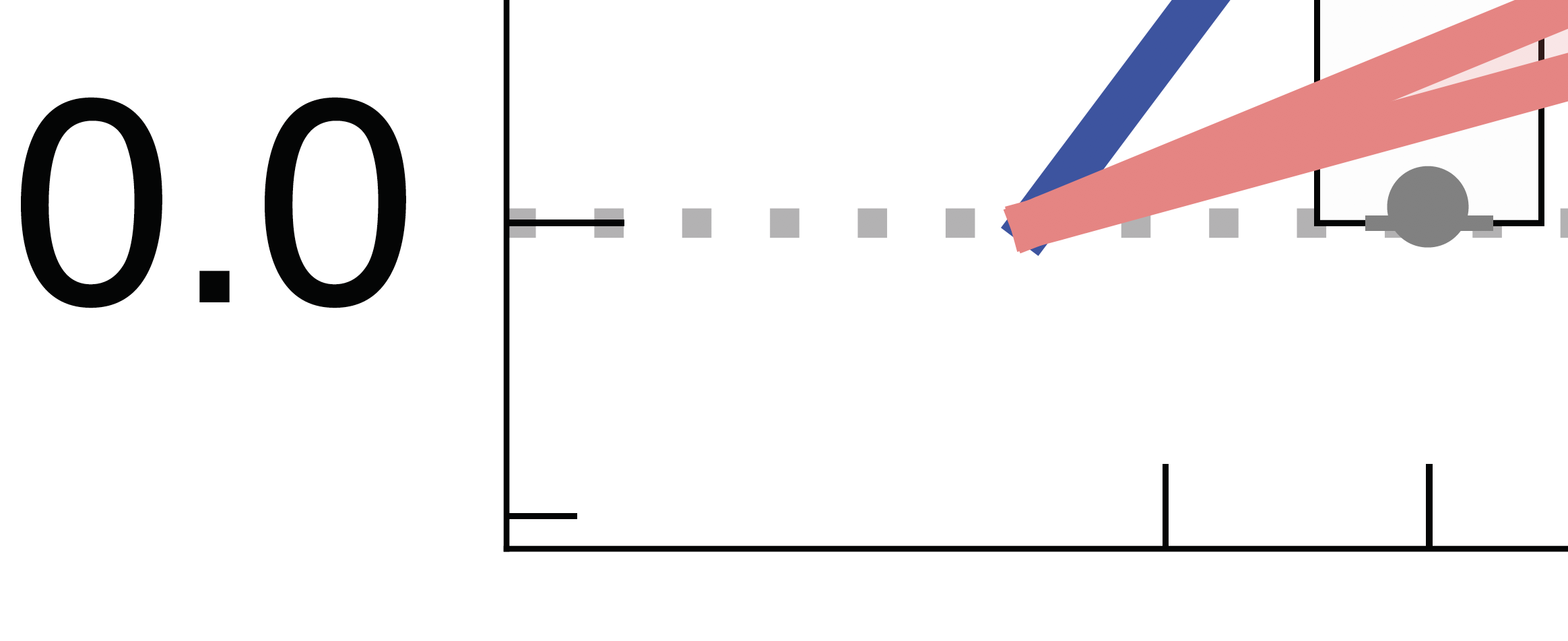}\\
\includegraphics[width=12cm]{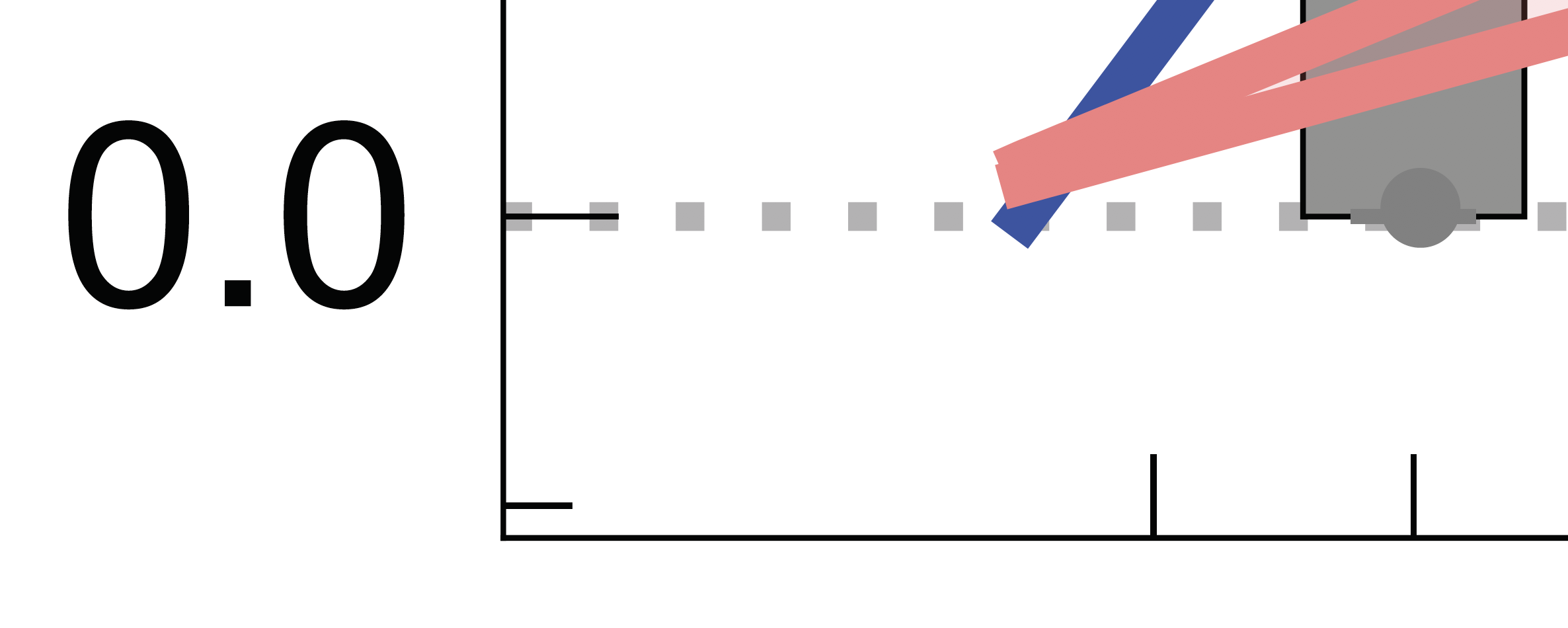}

    \end{center}
    \caption{Enlargement of Figs.\! 4e (top) and 4f (bottom) near the origin.
    We can see clearly that the red lines do not meet at the origin in Fig.\! 4f. 
    We have checked that the red lines meet at the origin in Figs.\! 4a--4d, similarly to Fig.\! 4e.
}
  \label{fig:fig4ef-origin}
\end{figure}

\end{enumerate}

In conclusion, the analysis of the paper with respect to Table 1 2014 Q3 and Figure 4f are not made using correct data and we cannot obtain any meaningful information from the table and figure.
The period 2014 Q3 is also
included in Fig.\! 5 of the second paper of the series by Miyazaki and Hayano~\cite{MH17}.  
If they used the same glass-badge data there, it is quite possible that also Figure 5 in Ref.~\cite{MH17} is made on the basis of, at least partially, false data and not reliable.
We also believe that, given the numerous anomalies pointed out in this Letter and in previous one \cite{OHKKMT20},
the authors have a duty of disclosing the entire software/codes/scripts used for the papers.

\subsubsection*{Acknowledgements}

We thank Ms. Akemi Shima for providing us with the public documents obtained through her Freedom of Information requests.
We also thank the \textit{KAGAKU} Editorial Office (Iwanami Shoten, Publishers) for opportunities to discuss this work.

\subsubsection*{Conflicts of interest}
Y.T.'s employment until February 2020 was funded through Programma per giovani ricercatori, anno 2014
``Rita Levi Montalcini'' of the Italian Ministry of Education, University and Research (MIUR).

\end{document}